\documentclass[aps,prd,showpacs,showkeys,superscriptaddress,nofootinbib]{revtex4-2}

\usepackage{epsfig,graphicx,amsfonts,amsmath,amssymb,bbm}
\usepackage{bbold}
\RequirePackage{mathrsfs}
\usepackage{color}
\usepackage{epsf}
\usepackage{epstopdf}

\def\eq#1{{Eq.~(\ref{#1})}}
\newcommand{\Le}{\left(}
\newcommand{\Ra}{\right)}

\newcommand{\beq}{\begin{equation}}
	\newcommand{\eeq}{\end{equation}}
\newcommand{\beqar}{\begin{eqnarray}}
	\newcommand{\eeqar}{\end{eqnarray}}
\newcommand{\D}{\partial}

\begin{document}
	\title{Time kink: modeling change of metric signature}
	\author{S. Bondarenko}
	\author{V. De La Hoz-Coronell}
	\affiliation{Ariel University, Ariel 4070000, Israel}

	\date{\today}
	
	\begin{abstract}
	
	The model of  a signature change of a metric from the Lorenztian to Euclidean one with the use of a time dependent kink as $g_{00}$ component of the metric is considered.
The metric which describes the continuous change of the signature of this type on a hypersurface is constructed and corresponding Einstein equations are solved in both regions of the space-time. The 
discontinuities of the Einstein tensor components on the hypersurface are discussed as well as junction conditions for the parameters of the 
solutions.
Additionally, the properties of a transition  from the space-time with one signature to an another are discussed, 
the presence of an inflation in the model
is demonstrated as a consequence of the signature change without any additional fields required.  
		
	\end{abstract}
	
	\maketitle

\section{Introduction}\label{Intro} 

 The different possible applications and constructions of the space time metric with changing signature is widely discussed in the literature, see many
aspects of the problem in \cite{MisWh,Hawk1,Sakh,Ander,Gero,Sork,Strom,Dray,Viss,Barv,borde,Kri,SigC,Green,Spont}. One from the possible manifestation of the Euclidean signature is a 
quantum world, see for example \cite{MisWh}. An another is a possible transition from the Euclidean regime to the Lorentizan one at the very beginning of the Universe,
see for example \cite{Hawk1}. There are different models of the signature change through mechanisms of the manifold complexification, non-flat tangent space and fields dynamics 
were proposed in the paper \cite{My} in the framework of the formalism of Einstein-Cartan gravity. 
An important component of many models proposed is a complex metric, see aslo discussions on the different aspects of the subject in 
\cite{Turok,Moffat,Penrose,Plebanski,Witten,Str,Volovich,Kon} for examples. usually, the models with the dynamical change of the signature are pretty involved mathematically, 
there are additional scalar or gauge field there which dynamics provides the signature transition. Therefore, it could be a good problem to consider a simple example with signature change which, hopefully, will clarify some general intrinsic 
properties of the signature change in general and in an application to the initial time cosmology in particular.

 Following to this idea, we consider the following construction. Let's introduce the $g_{00}$ component of the metric with argument as some hypersurface $\Psi(x)$, i.e. we define this metric's 
component as 
$g_{00}\,=\,g_{00}(\Psi(x))$. At next step will require that $g_{00}$ will have different signs for the positive and negative values of the argument, we wish 
\beq\label{Intro0}
g_{00}\,\xrightarrow[t\rightarrow \pm \infty]{\,}\,\pm 1
\eeq 
behavior of the metric's component will hold with
the property $g_{00}(\Psi(x)\,=\,0)\,=\,0$ satisfied.
Resolving the equation for the hypersurface and introducing a new variable $t\,=\,x^{0}\,-\,\Phi(\vec{x})\,=\,\Psi(x^0\,,\,x^i)$ we will arrive to the function $g_{00}(t)$ which is similar to a
time dependent kink, the model with  the $g_{00}(t)$ as the kink is a mostly natural choice therefore\footnote{There are models of gravity with kink-like solutions, see \cite{Fink} for example, where the space-time is rotated as whole in some two-dimensional plane, this is not a case we discuss.}.
There are also the following following important properties of the model  introduced  ab initio. First of all, we consider the continuous signature transition provided by continuous metric well defined at vicinity of $t\,=\,0$.
There is also a possibility that the signature change  can be happen similarly to the first order phase transition, we do not consider this case. An another important property of the model is that 
the contravariant metric component $g^{00}$ acquires a divergence at $t\,=\,0$ by definition of the metric, therefore a corresponding Einstein tensor and solution of the Einstein equations at vicinity
of $t\,=\,0$ has this induced singularity. We note also, that this singularity can be avoided in the models with complex metric where the change of the signature can
be achieved by the change of the phase of some complex quantity, see discussions concern different
aspects of the problem in \cite{My,Turok,Moffat,Penrose,Plebanski,Witten,Str,Volovich,Kon}, we do not consider this case and a complexification of the metric in this article.

 Basing on this construction, we would like to clarify the following questions. The first one is a properties and structure of the divergences arising in the Einstein tensor due the divergence
of the $g^{0 0}$. Calculating the tensor we, consequently, will face the following decomposition of it in both regions of the space-time:
\beq\label{Intro1}	
G_{\mu \nu}\,=\,G_{\mu nu}^{C}\,+\,G_{\mu \nu}^{D}\,
\eeq
with $G_{\mu \nu}^{C,D}$ as continuous and discontinuous parts correspondingly. The structure of the $G_{\mu \nu}^{D}$, in turn, will determine the structure of the corresponding discontinuous part of the energy-momentum tensor
of the separating hypersurface. Namely, the solution of the Einstein equations is possible in this model only if we will request the same form of the divergent parts in the both sides of the Einstein equations. Therefore, the knowledge of the behavior of the Einstein tensor  due the transition will provide the junction conditions for the parameters of the energy-momentum tensor in both space-time regions.
Technically it means that we find the local solutions of the Einstein equations at vicinity of $t\,=\,0$ in the model and basing on it's structure calculate the discontinuity of the Einstein tensor across the hypersurface that, in turn, determine the structure of the momentum-energy tensor. This procedure, of course, is similar to the usual junction conditions in the Einstein gravity, see 
for example \cite{Junc,Grav,Grav1}. The main difference, nevertheless, is that in the present model we do not consider delta-function like discontinuity but instead a discontinuity caused by the transition of 
$g_{00}$ through zero.
  
 An another interesting problem we can discuss is a time dependence of the Hubble parameter and inflating parameters in this modified cosmology. 
The desription of the cosmology is standard of course,  except that it describes the evolution of the Lorentzian space-time through it's transition from some Euclidean space-time region.
The calculation of the interesting parameters parameters in the model are usual, we introduce an observer's time $\tau$ in the model as
\beq\label{Intro2}
g_{00}(t)\,dt^{2}\,=\,d\tau^2\,
\eeq
that can be applied separately in the both patches of the whole space-time.
The main and direct consequence of the proposed transition between the regions and corresponding signature change we obtain, it is a presence of a geometrical inflation arises as a result of the dynamics of the transition.
Namely, solving the Einstein equation from both sides of the separating hypersurface we technically calculate the transition vertices between the Lorentzian and Euclidean regions. The solutions of course are dynamical ones, 
there is a time dependence of the vertices which is determined by the initial data of the Einstein equations. Depending on the vertices structure, therefore, we obtain an inflation and/or collapsing solutions for both regions, the vertices describe the combination of them from the both sides of the hypersurface . Important, that these inflation or shrinking of the space-times are geometrical ones, it is Einstein equations initial data that
determines the parameters of the inflation of collapsing and there are no any other fields are required.

  Respectively, the paper is organized as following. In the next Section we construct the metric of the problem and clarify it's main properties. The Section \ref{Ein} is dedicated to the calculation
of the Einstein tensor and corresponding local solutions at vicinity of the separating hypersurface. In this Section we also determine the structure of the energy-momentum tensor and junction conditions for the parameters of the solutions from the both regions of the space-time. In the Section \ref{Cosm} we investigate a inflation properties of the Lorentzian space-time arose from the transition from an Euclidean region and discuss corresponding vertices of the transition. 
The last Section is a Conclusion where we discuss the main results of the paper.

\section{Modeling signature change}\label{Sign} 

 In the formalism which describes signature transition, we introduce ad hoc the following specially constructed metric:
\beq\label{Kin1}
ds^{2}\,=\,\varphi(\Psi(x))\,dx^{0 2}\,-\,2\,\tilde{\sigma}_{i 0}(x^{0},x)\,dx^{0}\,dx^{i}\,-\,
\sigma_{i j}(x^{0},x)\,dx^{i}\,dx^{j}\,,
\eeq
with 
\beq\label{Kin2}
\varphi(x)\,=\,\tanh(\lambda\,\Psi(x))
\eeq
as time-dependent kink with with parameter $\lambda$.
Now, in order to 
describe the dynamics of the kink, we consider a hypersurface of the separation given by 
\beq\label{Kin3}
\Psi(x^0\,,\,x^i)\,=\,x^{0}\,-\,\Phi(\vec{x})\,=\,0\,.
\eeq
With the new  variable
\beq\label{Kin4}
t\,=\,x^{0}\,-\,\Phi(\vec{x})\,
\eeq
 the \eq{Kin1} metric reduces to
\beq\label{Kin5}
ds^{2}\,=\,\varphi(t)\,dt^{2}\,+\,2\,\Le \varphi\,\Phi_{,i} - \tilde{\sigma}_{i 0}\Ra\,dx^{i}\,dt\,+\,
\Le \varphi\,\Phi_{,i}\,\Phi_{,j}\,-\,2\,\tilde{\sigma}_{i 0}\,\Phi_{,j}\,-\,\sigma_{i j}\Ra\,dx^{i}\,dx^{j}\,.
\eeq
Defining
\beq\label{Kin6}
\varphi\,\Phi_{,i} \,=\, \tilde{\sigma}_{i 0}
\eeq
we obtain for the metric
\beq\label{Kin7}
ds^{2}\,=\,\varphi(t)\,dt^{2}\,-\,\Le \sigma_{i j}\,+\,\varphi\,\Phi_{,i}\,\Phi_{,j}\Ra\,dx^{i}\,dx^{j}\,=\,
\varphi(t)\,dt^{2}\,-\,h_{i j}\,dx^{i}\,dx^{j}\,.
\eeq
Here the induced metric
\beq\label{Kin8}
h_{ij}(x)\,=\,g_{\mu \nu}\,\frac{\D x^{\mu}}{\D y_{i}}\,\frac{\D x^{\nu}}{\D y_{j}}\,=\,\Le \varphi(t)\,\Phi_{,i}\,\Phi_{,j}\,+\,\sigma_{i j} \Ra\,
\eeq
on the hypersurface 
\beq\label{Kin9}
t\,=\,0\,
\eeq
is introduced with
$y^{i}\,=\,x^{i}\,,i=1\dots 3 $ parametrization of the hypersurface. 
Further we assume that the condition
$\Le \varphi(-t)\,\Phi_{,i}\,\Phi_{,j}\,+\,\sigma_{i j}(-t)\Ra\,>\,0$ is satisfied in order of preserving the Euclidean signature at $t\,<\,0$.
The vecor normal to the space-like surface can be found as well, it is equal to
\beq\label{Kin10}
n_{\mu}\,=\,\frac{t_{,\mu}}{\sqrt{g^{\nu \rho} t_{,\nu} t_{,\rho}}}\,.
\eeq
Using the inverse metric
\beq\label{Kin11}
g^{\mu \nu}\, =\, 
\left(
\begin{array}{cc}
\varphi^{-1} & 0 \\
0 & -\,\Le \sigma_{i j}\,+\,\varphi\,\Phi_{,i}\,\Phi_{,j}\Ra^{-1}
\end{array} \right)\,=\,
\left(
\begin{array}{cc}
\varphi^{-1} & 0 \\
0 & -\,\tilde{h}^{i j}
\end{array} \right)\,
\xrightarrow[t\rightarrow 0]{\,}\,
\left(
\begin{array}{cc}
\varphi^{-1} & 0 \\
0 & -\sigma_{i j}^{-1}
\end{array} \right)
\eeq
we obtain for the normal covariant vector 
\beq\label{Kin11001}
n_{\mu}\,=\,\varphi^{1/2}(t)\,\Le 1,0,0,0 \Ra\,
\eeq
and for the contrvariant one correspondingly:
\beq\label{Kin11002}
n^{\mu}\,=\,\varphi^{-1/2}(t)\,\Le 1,0,0,0 \Ra\,
\eeq
that, definitely, can be immediately obtained from \eq{Kin7} expression. The \eq{Kin7} metric describes the space-time with Lorentzian signature at $t\,>\,0$ and with Euclidean one at $t\,<\,0$ and,
as expected, the $g^{0 0}$ metric's component and corresponding $n^{\mu}$ have discontinuity 
in vicinity of $t\,=\,0$,  we assumed here also that the $\sigma_{i j}(t\,=\,0)\,\neq \,0$ and has no discontinuity at $t\,=\,0$\,.

\section{Einstein equations at vicinity of transition boundary}\label{Ein} 

 In order to clarify the discontinuities of the Einstein equations caused by \eq{Kin11} metric we calculate a Ricci tensor for the model, we have:
\beqar\label{DW3}
R_{00} & = &-\,\frac{1}{2}\frac{\D \varkappa^{i}_{i}}{\D t}\,-\,\frac{1}{4}\,\,\varkappa^{i}_{j}\,\varkappa^{j}_{i}\,
+\,\frac{1}{4} g^{00}\,\frac{\D g_{00}}{\D t}\,\varkappa^{i}_{i}\,,\nonumber\\
R_{0i} & = & \frac{1}{2} \Le \varkappa^{j}_{i;j}\,-\,\varkappa^{j}_{j;i} \Ra\,,\nonumber \label{DW301} \\
R_{i j}& = & P_{i j}\,+\,\frac{1}{2}\,g^{00}\,\frac{\D \varkappa_{i j}}{\D t}\,
+\,\frac{1}{2}\,\frac{\D g^{00}}{\D t}\,\varkappa_{i j}\,+\,\frac{1}{4}\,\Le g^{00}\Ra^{2}\,\frac{\D g_{00}}{\D t}\,\varkappa_{i j}\,
+\,\frac{1}{4}\,g^{0 0}\, \Le \varkappa_{ij}\,\varkappa^{k}_{k}\,-\,2\,
\varkappa^{k}_{i}\,\varkappa_{j k} \Ra\,\label{DW302}.
\eeqar
with 
\beq\label{DW4}
\varkappa_{i j}\,=\,\frac{\D h_{i j}}{\D t }
\eeq
and $P_{ij}$ as 3D Ricci tensor.
Correspondingly, we can write the Einstein tensor in the following form isolating separately continuous ($G_{00}^{C}$) and potentially discontinuous parts:
\beqar\label{DW31}
G_{00} & = & R_{00}\,-\,\frac{1}{2}\,g_{00}\,\Le g^{00}R_{00}\,+\,g^{i j} R_{i j} \Ra\,=\,\frac{1}{2}\,R_{00}\,+\,\frac{1}{2}\,g_{0 0}\,\tilde{h}^{i j}\,R_{i j} \,=\, \nonumber \\
&=&\,\frac{1}{4}\,g^{00}\,\frac{\D g_{00}}{\D t}\,\varkappa^{i}_{i}\,+\,\frac{1}{4}\,g_{00}\,\frac{\D g^{00}}{\D t}\,\tilde{h}^{i j}\,\varkappa_{i j}\,+\,
G^{C}_{00}\,=\,\frac{1}{4}\,\frac{\D \Le g^{00}\,g_{00}\Ra}{\D t}\,\varkappa^{i}_{i}\,+\,G^{C}_{00}\,=\,G^{C}_{00}\,.
\eeqar
For the mixed tensor's components we obtain
\beq\label{DW33}
G_{0 i}\,=\,R_{0 i}\,,
\eeq
and for the spatial tensor's components we have
\beqar\label{DW34}
G_{ij } & = &\frac{1}{2}\,g^{00}\,\Le \frac{\D \varkappa_{i j}}{\D t}\,-\,h_{i j}\,\frac{\D \varkappa_{ l}^{l}}{\D t}\, +\,
\frac{1}{2}\,\Le \varkappa_{i j}\,\varkappa^{l}_{l}\,-\,2\,\varkappa^{l}_{i}\,\varkappa_{j l}\Ra\,-\,
\frac{1}{4}\,h_{i j}\,\Le \varkappa_{k}^{k}\,\varkappa^{l}_{l}\,-\,\varkappa^{l}_{k}\,\varkappa_{l}^{k}\Ra\,\Ra\,+\,\nonumber \\
&+&\,
\frac{1}{2}\,\frac{\D g^{00}}{\D t}\,\Le \varkappa_{i j}\,-\,\frac{1}{2}\,h_{i j}\,\varkappa_{l}^{l}\Ra\,+\,
\frac{1}{4}\,( g^{00})^{2}\,\frac{\D g_{00}}{\D t}\,\varkappa_{i j}\,+\,G_{i j}^{C}\,=\,G_{i j}^{D}\,+\,G_{i j}^{C}\,.
\eeqar
Using an expansion of the exponential of \eq{Kin2} function at $t\,\rightarrow \delta\,$ as
\beq\label{DW5}
e^{\lambda\, t}\,\rightarrow\,1\,+\,\lambda\,\delta\,,
\eeq 
we can calculate the discontinuous part of the $G_{i j}$ tensor obtaining the following expression:
\beqar\label{DW6}
G_{ij}^{D} & = &\frac{1}{2}\,\frac{1}{\lambda \,\delta}\,\Le \frac{\D \varkappa_{i j}}{\D t}\,-\,h_{i j}\,\frac{\D \varkappa_{ l}^{l}}{\D t}\, +\,
\frac{1}{2}\,\Le \varkappa_{i j}\,\varkappa^{l}_{l}\,-\,2\,\varkappa^{l}_{i}\,\varkappa_{j l}\Ra\,-\,
\frac{1}{4}\,h_{i j}\,\Le \varkappa_{k}^{k}\,\varkappa^{l}_{l}\,-\,\varkappa^{l}_{k}\,\varkappa_{l}^{k}\Ra\,\Ra\,-\,\nonumber \\
&-&\,
\frac{1}{2}\,\frac{1}{\lambda \,\delta^{2}}\,\Le \varkappa_{i j}\,-\,\frac{1}{2}\,h_{i j}\,\varkappa_{l}^{l}\Ra\,+\,
\frac{1}{4}\,\frac{1}{\lambda \,\delta^2}\,\varkappa_{i j}\,+\,G_{i j}^{C}\,=\,G_{i j}^{D}\,+\,G_{i j}^{C}\,.
\eeqar
The leading, quadratically divergent, parts of the expression does not contribute into the discontinuity and we obtain
\beq\label{DW8}
G_{i j}^{D}\,=\,-\,\frac{1}{4\,\lambda\,\delta^{2}}\,\Le \varkappa_{i j}\,-\,h_{i j}\,\varkappa_{l}^{l}\Ra\,.
\eeq
Now we can match the Einstein tensor with energy-momentum tensor $T_{\mu \nu}$ determining it's discontinuous parts separately in the Lorentzian and Euclidean regions.

   We determine the energy-momentum tensor in both regions, Euclidean (E) and Lorentzian one (L), as sum of it's continuous and discontinuous parts:
\beq\label{DW9}
T_{\mu \nu}^{L,E}\,=\,T_{\mu \nu}^{(L,E)C}\,+\,T_{\mu \nu}^{(L,E) D}\,,
\eeq
further we will consider only $T_{\mu \nu}^{D}$ part of the stress-energy tensor and corresponding parts of the Einstein tensor. 
We, also, look for the solution of the equations in the vicinity of the $\delta\,\rightarrow\,0$ point, therefore
it will be enough to preserve in the corresponding equations only mostly divergent parts.
Now, first of all, we consider the Lorentzian sector of the space time, we define the steress-energy tensor in the form which corresponds to the found divergences:
\beqar\label{DW10}
T_{0 0}\,^{L D}\,& = &\,0\,\nonumber \\
T_{0 j}\,^{L D}\,& = &\,0\,, \nonumber \\
T_{i j}\,^{L D}\,& = &\,\frac{1}{\delta^2}\,p^{L}\,h_{i j}\,.
\eeqar
In order to solve the Einstein equation consists \eq{DW8} and corresponding r.h.s. from \eq{DW10}
\beq\label{DW11}
-\,\frac{1}{4\,\lambda\,\delta^{2}}\,\Le \varkappa_{i j}\,-\,h_{i j}\,\varkappa_{l}^{l}\Ra\,=\,\frac{1}{\delta^2}\,p^{L}\,h_{i j}\,
\eeq
we introduce the following ansatz valid in teh Lorentzian and Euclidean parts of the space-time:
\beqar\label{DW111}
&\,& h_{i j}\,=\,h_{0 i j}\,e^{-\,\mathcal{E} t}\,,\,\,\,\tilde{h}^{i j}\,=\,\tilde{h}^{i j}_{0}\,e^{\,\mathcal{E} t}\, \\
&\,&h_{i j}\,\tilde{h}^{j k}\,=\,\delta^{k}_{i} \,,\,\,\, h_{0 l}^{\,l}\,=\,C\,.
\eeqar
The \eq{DW11} has no spatial derivative, in general the $C,\mathcal{E}$ and $p^{L}$ parameters can be arbitrary functions of spatial coordinates, but for the sake of simplicity we assume that the parameters are constants. So, we obtain for the \eq{DW11} in the Lorentzian sector:
\beq\label{DW12}
\mathcal{E}\,h_{0 i j}\,e^{-\,\mathcal{E} t}\,\Le 1\,-\,C  \Ra\,=\,4\,\lambda\,p^{L}\,h_{0 i j}\,e^{-\,\mathcal{E} t}\,
\eeq	
or
\beq\label{DW13}
C\,=\,1\,-\frac{\,4\,\lambda\,p^{L}}{\mathcal{E}}\,.
\eeq
Consequently, we can use an another Einstein equation with $G^{C}_{00}$ from \eq{DW31} and corresponding energy-momentum tensor in order to relate $p^{L}$ and $\mathcal{E}$
parameters, we have there:
\beq\label{DW14}
\frac{1}{8}\,\Le \varkappa_{l}^{l}\,\varkappa_{k}^{ k}\,-\,3\,\varkappa_{l}^{k}\,\varkappa_{k}^{l}\,\Ra\,=\,T_{00}^{LC}\,=\,\epsilon^{L}.
\eeq
Inserting \eq{DW111} ansatz in this equation we obtain correspondingly the equation for the $\mathcal{E}$ in terms of $p^{L}$ and $\epsilon^{L}$:
\beq\label{DW15}
\mathcal{E}^{2}\,\Le C^{2}\,-\,9\,\Ra\,=\,8\,\epsilon^{L}\,
\eeq
that provides two different solutions:
\beqar\label{DW16} 
&\,& \mathcal{E}_{1}^{L}\,=\,-\frac{1}{2}\,\Le \lambda\,p^{L}\,-\,\sqrt{9(\lambda\,p^{L}\,)^2\,-\,4\,\epsilon^{L}}\Ra \,,\,\,\,C_{1}^{L}\,=\,1\,-\,
\frac{4\,\lambda\,p^{L}}{\mathcal{E}_{1}^{L}}\,,\\
&\,& \mathcal{E}_{2}^{L}\,=\,-\frac{1}{2}\,\Le \lambda\,p^{L}\,+\,\sqrt{9(\lambda\,p^{L}\,)^2\,-\,4\,\epsilon^{L}}\Ra \,,\,\,\,C_{2}^{L}\,=\,1\,-\,
\frac{4\,\lambda\,p^{L}}{\mathcal{E}_{2}^{L}}\,\,\label{DW1601} .
\eeqar
The last Einstein equation, $G_{i 0}\,=\,0\,$, is satisfied identically for the \eq{DW111} ansatz in the case of the constant values of the parameters.
We also note, that if we want preserve the Lorentzian signature of the \eq{Kin7} metric, we have to consider in the the \eq{DW16}-\eq{DW1601} only the values of the parameters which provide
positive $h_{i j}$ metric's components.

 For the Euclidean part of space time we also define the energy-momentum tensor as
\beqar\label{DW17}
T_{0 0}\,^{E D}\,& = &\,0 \\
T_{0 j}\,^{E D}\,& = &\,0\,\nonumber \\
T_{i j}\,^{E D}\,& = &\,\frac{1}{\delta^2}\,p^{E}\,h_{i j}\,.  
\eeqar
Due the squared $\delta$ in the divergence part of the equations, we will obtain, therefore, that for the Euclidean sector we have the same answers as for the Lorenzrian one, with only corresponding change of the parameters. Also, because of $[G_{00}]\,=\,0$, we have
\beq\label{DW173}
\epsilon^{L}\,=\,\epsilon^{E}\,.
\eeq
Therefore, at $t\rightarrow\,0$ limit, we can consider two different limiting cases :
\beqar\label{DW18}
&\,& p^{E}\, = \,p^{L}\,,\\
&\,& [G_{i j}]\,\stackrel{t\rightarrow 0}{=}\,\,p^{L}\Le h_{0 ij}^{L}\,-\,h_{0 ij}^{E}\Ra\,,\label{DW181} \\
&\,&\mathcal{E}_{1 2}^{E}\,=\,-\frac{1}{2}\,\Le \lambda\,p^{E}\,\mp\,\sqrt{9(\lambda\,p^{E}\,)^2\,-\,4\,\epsilon^{E}}\Ra \,=\,\mathcal{E}_{1 2}^{L} \,,\,\,\,C_{12}^{E}\,=\,1\,-\,
\frac{4\,\lambda\,p^{E}}{\mathcal{E}_{12}^{E}}\,=\,C_{12}^{L}\,\label{DW182} 
\eeqar
and
\beqar\label{DW19}
&\,& p^{E}\, = \,-\,p^{L}\,,\\
&\,& [G_{i j}]\,\stackrel{t\rightarrow 0}{=}\,\,p^{L}\Le h_{0 ij}^{L}\,+\,h_{0 ij}^{E}\Ra\,,\label{DW191} \\
&\,& \mathcal{E}_{12}^{E}\,=\,\frac{1}{2}\,\Le \lambda\,p^{L}\,\pm\,\sqrt{9(\lambda\,p^{L}\,)^2\,-\,4\,\epsilon^{L}}\Ra \,,\,\,\,C_{12}^{E}\,=\,1\,+\,
\frac{4\,\lambda\,p^{L}}{\mathcal{E}_{1 2}^{E}}\,.\,\label{DW192}
\eeqar
Again, from the all possible values of the  parameters, we have to keep only these which will preserve the Euclidean signature of the initial metric, i.e. positive values of $h_{i j}$.
The obtained results we will discuss in the Conclusion of the paper.


\section{Vertices of cosmological transition and an inflation as result of the signature change}\label{Cosm} 

 In the given context, the solution of Einstein equations for the given metric means a calculation of the spatial components of the metric at given $g_{00}$ component, see \eq{Kin7}. Due the not-trivial
form of the metric only at vicinity $t\,-\,\Psi(\vec{x})\,=\,0$, we are looking for a local form of the solution for the  Lorentzian space-time and corresponding Euclidean space-time around the transition hypersurface. The model describes a transition of the one part of the space-time to another in different forms depending on the time's variable arrow. Further we assume the direction of the
evolution of the Lorentzian or Euclidean space-times basing only on the time behavior of the initial hypersurface at initial moment of time. The positive sign of the time
variable defines a positive arrow and the negative sign correspondingly denotes the negative direction. In this case the possible solutions determine types of the transitions vertices between different geometries with different times directions, also their form determine the evolution scenario for the both manifolds.

  In order to analyze the inflation properties of the evolutioning space-time we proceed as usual.
The obtained model describes an  updated regular Friedman isotropic model with the metric given by the following expression:
\beq\label{cm1}
ds^{2}\,=\,
\varphi(t)\,dt^{2}\,-\,a^{2}(t)\,dl^2\,=\,d\tau^{2}\,-\,a^{2}(\tau)\,dl^2\,
\eeq
with $d^2 l$ as usual mostly symmetric volume element of the three-dimensional space. Here we introduce the proper time via
\beq\label{cm11}
d\tau\,=\,\sqrt{\varphi(t)}\,dt\,.
\eeq
We note, that the $t$ time can not be expressed through  $\tau$ in the both regions of the space-time because the proper time integral for the $\tau$
\beq\label{cm111}
\tau\,=\,\int^{t}\,\sqrt{\varphi(t)}\,dt
\eeq
can not be continued at the region of the negative $t$, i.e it is not defined in the Euclidean region of the space-time for negative $t$ with 
$\varphi(t)\,=\,\tanh(\lambda\,t)$.  Therefore, the \eq{cm11} definition of the proper time is valid only in the Lorentzian path of the whole space-time.
We have for the \eq{cm111} in the case of the Lorentzian space-time:
\beq\label{cm11101}
\tau\,=\,\int^{t}\,\sqrt{\tanh(\lambda\,t)}\,dt\,\approx\,\frac{2}{3}\,\sqrt{\lambda}\,t^{3/2}
\eeq
valid at vicinity of $t\,=\,0$.
Further we rewrite the \eq{cm1} obtaining
\beq\label{cm112}
ds^{2}\,=\,\,a^{2}(\eta)\,\Le 
d\eta^{2}\,-\,dl^2\Ra\,
\eeq
with regulat additional redifinition of the proper time variable. Assuming that in \eq{Kin7} we have $h_{ij}\,=\,h\delta_{i j}$ , we obtain that 
\beq\label{cm113}
a^{2}(\eta)\,=\,h(\eta)\,,\,\,\,a(\eta)\,=\,\pm\,\sqrt{h(\eta)}
\eeq
with $h^{L,E}$ in the form obtained in the previous Section and reduced to the diagonal form. The last intermediate step we need is to relate the $t$ and $\eta$ variables,for that we will calculate
\beq\label{cm114}
\phi^{1/2}(t)\, h^{-1/2}(t)\,dt\,=\,d\eta
\eeq
for the each $h$ separately. Correspondingly, considering the Hubble constnat through it's usual definition in terms of the conformal time $\eta$:
\beq\label{cm06}
H\,=\,\frac{d a(\eta)/d \eta}{a^{2}(\eta)}\,=\,h^{-1}\frac{d a(\tau)}{d \tau}\,\frac{d \tau}{d t}\,\frac{d t}{d \eta}\,=\,h^{-1/2}(\tau)\frac{d \sqrt{h(\tau)}}{d \tau}\,=\,\frac{1}{2}\frac{d \ln h}{d \tau}\,,
\eeq
here the positive root of \eq{cm113} was used.
Of course the expression is valid in only very first moments of a space-time evolution, the inflation (or anti-inflation) type of the evolution is determined by the \eq{cm113} identity which determines the time dependence
of $a$ factor.

 Now, first of all, we consider the \eq{DW16}-\eq{DW1601} solutions for the Lorentiazn space-time, we will take only positive $p^{L}$ allowing to $p^{E}$ and $\epsilon^{L}$ to be positive and negative.
\begin{enumerate}
\item 
Consider the solution with $C\,>\,0,\,$ $\epsilon^{L}\,>\,0$. For the simplicity we also will avoid the complex solutions taking the following equation of state
\beq\label{Conc1}
9\,\Le \lambda\,p^{L}\,\Ra^2\,=\,4\,\epsilon^{L}\,.
\eeq 	
We note, that the $p^{L}$ and $\epsilon^{L}$	belong to the discontinuous and continuous parts of the energy-momentum tensor and therefore have different dimensions in general. 
Now we obtain: 
\beq\label{Conc2}	
\mathcal{E}_{1}^{L}\,=\,\mathcal{E}_{2}^{L}\,=\,-\,\frac{1}{3}\,\sqrt{\epsilon^{L}}\,,C\,=\,9\,
\eeq	
and consequently
\beq\label{Conc3}	
h_{i j}^{L}\,=\,h_{0 i j}^{L}\,e^{\,\frac{1}{3}\,\sqrt{\epsilon^{L}} t}\,.
\eeq
We have here a solution which growing with time at least locally, i.e. expanding solution. Further we take $h_{0 i j}=\,h_{0}\,\delta_{i j}$ and obtain through the \eq{cm113} and \eq{cm06}:
\beq\label{Conc301}	
a(\tau)\,=\,h^{1/2}_{0}\,e^{\,\frac{\sqrt{\epsilon^{L}}}{3\,\lambda^{1/3}}\, \Le 3\tau / 2\Ra^{2/3}}\,=\,h^{1/2}_{0}\,e^{\,\frac{\sqrt{T_{00}^{L}}}{3\,\lambda^{1/3}}\, \Le 3\tau / 2\Ra^{2/3}}\,
\eeq
and 
\beq\label{Conc302}	
H(\tau)\,\propto\,\Le \frac{2}{3}\Ra^{1/3}\,\frac{\sqrt{T_{00}^{L}}}{3\,(\lambda\,\tau)^{1/3}}\, 
\eeq
which describe a scenario with exponentially growing $a(\tau)$, i.e. inflation scenario, and with Hubble constant which has singularity at the hypersurface of separation between Lorentzian and Euclidean sectors.

\item
 Now we request $C\,>\,0,\,$ taking $\epsilon^{L}\,<\,0$, the negative energy density arises in the inflation scenario,  see \cite{LindeInfl} for example. 
We have for the \eq{DW16} case:
\beq\label{Conc4}	
\mathcal{E}_{1}^{L}\,=\,-\frac{1}{2}\,\Le \lambda\,p^{L}\,-\,\sqrt{9(\lambda\,p^{L}\,)^2\,+\,4\,|\epsilon^{L}|}\Ra \,>\,0
\eeq
with two possibilities
\beq\label{Conc5}	
\mathcal{E}_{1\,a}^{L}\,>\,4\,\lambda\,p^{L}\,,\,\,\,\mathcal{E}_{1\,b}^{L}\,<\,4\,\lambda\,p^{L}\,
\eeq
where for both cases the condition
\beq\label{Conc6}
C\,>\,0
\eeq
must be satisfied.In this case we have 	
\beq\label{Conc7}	
h_{i j}^{L}\,=\,h_{0 i j}^{L}\,e^{\,-\,|\mathcal{E}_{1}^{L}| t}\,,
\eeq
i.e. a solution for the shrinking spatial metric. In this case we obtain correspondingly:
\beq\label{Conc701}	
a(\tau)\,=\,h^{1/2}_{0}\,e^{\,-\,\frac{|\mathcal{E}_{1}^{L}|}{\lambda^{1/3}}\, \Le 3\tau / 2\Ra^{2/3}}\,
\eeq
and 
\beq\label{Conc702}	
H(\tau)\,\propto\,-\,\Le \frac{2}{3}\Ra^{1/3}\,\frac{|\mathcal{E}_{1}^{L}|}{(\lambda\,\tau)^{1/3}}\, 
\eeq
that describes collapsing Lorentzian space-time, i.e. anti-inflation scenario.

\item
For the negative $\epsilon^{L}\,<\,0$ there is also \eq{DW1601} solution which provides:
\beq\label{Conc8}	
\mathcal{E}_{2}^{L}\,=\,-\frac{1}{2}\,\Le \lambda\,p^{L}\,+\,\sqrt{9(\lambda\,p^{L}\,)^2\,+\,4\,|\epsilon^{L}|}\Ra \,<\,0
\eeq
and
\beq\label{Conc9}	
h_{i j}^{L}\,=\,h_{0 i j}^{L}\,e^{\,|\mathcal{E}_{2}^{L}| t}\,,
\eeq
as expanding solution. Therefore we obtain:
\beq\label{Conc70101}	
a(\tau)\,=\,h^{1/2}_{0}\,e^{\,\frac{|\mathcal{E}_{2}^{L}|}{\lambda^{1/3}}\, \Le 3\tau / 2\Ra^{2/3}}\,
\eeq
and 
\beq\label{Conc70202}	
H(\tau)\,\propto\,\Le \frac{2}{3}\Ra^{1/3}\,\frac{|\mathcal{E}_{2}^{L}|}{(\lambda\,\tau)^{1/3}}\, 
\eeq
that describes the inflation scenario for the Lorentzian space-time as well.
\end{enumerate}

 The Euclidean space-time metric's solution we can separate on the two parts. The first one, \eq{DW18}-\eq{DW182}, we can call as symmetric counterpart of the corresponding Lorentzian solutions. 
The second part, \eq{DW19}-\eq{DW192}, are different from the Lorentizan ones. Correspondingly to the previously described Lorenztian metrics we have the following Euclidean counterparts in this case.
\begin{enumerate}
\item 
With the values of the parameters $C\,>\,0,\,$ $\epsilon^{L}\,>\,0$ and taking
\beq\label{Conc1001}
9\,\Le \lambda\,p^{L}\,\Ra^2\,=\,4\,\epsilon^{L}\,.
\eeq 	
we obtain: 
\beq\label{Conc2001}	
\mathcal{E}_{1}^{E}\,=\,\mathcal{E}_{2}^{E}\,=\,\frac{1}{3}\,\sqrt{\epsilon^{L}}\,,C\,=\,9\,
\eeq	
and consequently
\beq\label{Conc3001}	
h_{i j}^{E}\,=\,h_{0 i j}^{E}\,e^{\,-\,\frac{1}{3}\,\sqrt{\epsilon^{L}} t}\,,
\eeq
which is shrinking Euclidean solution.

\item
 Now we request $C\,>\,0,\,$ taking $\epsilon^{L}\,<\,0$, the negative energy density arises in the inflation scenario,  see \cite{LindeInfl} for example. 
We have for the \eq{DW16} case:
\beq\label{Conc4001}	
\mathcal{E}_{1}^{E}\,=\,\frac{1}{2}\,\Le \lambda\,p^{L}\,+\,\sqrt{9(\lambda\,p^{L}\,)^2\,+\,4\,|\epsilon^{L}|}\Ra \,>\,0
\eeq
with 
\beq\label{Conc6001}
C\,>\,0\,.
\eeq
In this case we have 	
\beq\label{Conc7001}	
h_{i j}^{E}\,=\,h_{0 i j}^{E}\,e^{\,-\,\mathcal{E}_{1}^{E} t}\,,
\eeq
i.e. again here there is a shrinking kind of the solution.

\item
For the negative $\epsilon^{L}\,<\,0$ there is also the second root of \eq{DW192} expression which provides:
\beq\label{Conc8001}	
\mathcal{E}_{2}^{E}\,=\,\frac{1}{2}\,\Le \lambda\,p^{L}\,-\,\sqrt{9(\lambda\,p^{L}\,)^2\,+\,4\,|\epsilon^{L}|}\Ra \,<\,0
\eeq
and
\beq\label{Conc9001}	
h_{i j}^{E}\,=\,h_{0 i j}^{E}\,e^{\,|\mathcal{E}_{2}^{E}| t}\,,
\eeq
as expanding solution. Here we request as well that the values of the parameter preserve $C\,>\,0$. 
\end{enumerate}
We do not discuss here the Hubble parameter for the Euclidean space-time, it is meaning is not really clean in the case of euclidean metric given by \eq{Kin7} for the case of negative time variable. Nevertheless, from teh results is clear that
similarly to the Lorentzian patch, here we also have the expanding or shrinking Euclidean solutions as counterparts of the Euclideant sector evolution.

 \section{Conclusion}\label{Fin} 

 In the article we introduced a simple model of a metric with signature change through the continuous change of the sign of the $g_{00}$ metric's  component.
Similar models can be formulated in the framework of the Einstein-Cartan gravity with dynamical vielbein fields and/or complexification of the space-time manifold, see  \cite{My} for examples.
Nevertheless, we found is useful to discuss firstly a simpler model of the signature transition.
The description of the possible consequences of such transition from the Euclidian to Lorentzian space-time in such a model  is a main purpose of the manuscript.

 The different ways of the separation of the whole space-time on the regions with different metrics and further 
matching of the metrics by the junction conditions it is a well known procedure in the 
general relativity theory of course, see \cite{Junc,Grav,Grav1,DWall} for examples. In these calculations the 
separation of the space-time and corresponding discontinuity was provided mainly by the delta-function defined on some hypersurface of separation. In our model we considered a different, local, approach to the problem
of the Euclid-Lorentz space time regions transition. From the very beginning we assume that the $g_{00}$ component of the metric is changing as a kink which depends on "time" variable.
This variable is defined through the separation hypersurface being zero on it. Therefore, we initially know the divergences arising in the geometry and 
in the all corresponding calculations due the
contravariant $g^{00}$ metric's component, i.e. it is due the transition of $g_{00}$ through the zero. As a result we can analyze the corresponding discontinuities of the Einstein tensor and Einstein equations expanding locally all coresponding equations at vicinity of $t\,=\,0$.

 So, in the given context, the solution of Einstein equations for the given metric means a calculation of the spatial components of the metric at given $g_{00}$ component, see \eq{Kin7}. Due the not-trivial
form of the metric only at vicinity $t\,-\,\Psi(\vec{x})\,=\,0$, it means that we are looking for a local form of the solution for the  Lorentzian space-time and corresponding Euclidean space-time around some special moment of time. In the present model this moment and arrow of the time can be any in fact, the singularity is proportional to the square of the $\Delta\,t$. That means that the model can describes a transition of the one part of the space time to another, their both creations or the end of both, depending on the assumed time's variable arrow. Therefore, we can assume the direction of the
evolution of the Lorentzian or Euclidean space-times basing only on the time behavior of the initial hypersurface at initial moment of time. In the calculation we accept that the positive sign of the time
variable defines a positive arrow and the negative sign correspondingly denotes the negative direction. In this case the possible solutions
of the Einstein equations determine some types of the transitions vertices between different geometries with possible different times directions from both sides of the hypersurface. Important to underline, that the obtained solutions are local
and do not determine the evolution of the manifold at general.
Therefore, the obtained results we can interpret as a calculation of some vertices of transition from Lorentzian to Euclidean and vice verse geometries.
As transition vertices in the model, of course, we can call the pairs of the $h_{ij}^{L,E}$ tensors which describes the evolution of the space-time from the different areas of the separating hypersurface.
Correspondingly, we obtained the all possible variants of the mutual evolution of the metric at the initial time moment in the different parts of the space-time. 
As clarified above, here the variants are the time directions of the metric evolution defined by the sign in front of the coefficient in the power of the corresponding exponential.
The natural 
description of the these kind of the transitions arise in the theory of the not flat tangent spaces, see \cite{My}, see also \cite{MyWorm}
where similar constructions for the different type of the transition vertices were discussed.

 An additional property of the proposed model is a presence of an inflation, and/or collapsing, behavior of the space-time regions 
simply as a consequence of the transition between the geometries. Considering the \eq{DW18}-\eq{DW19} junction conditions we see that the important parameters there are only the values of the initial energy-momentum tensor.
So we have the initial data for the Einstein equations, see \eq{DW10} and \eq{DW14} for the definitions of the initial energy and momentum of the corresponding energy-momentum tensor, as only parameters present in the model. 
Therefore, the calculated transition vertices
are depend on these two parameters only, there are no any additional fields are required in order to obtain exponentially growing, or collapsing, Universe. Interesting, that considering for example
\eq{Conc1} solution for the $h_{ij}^{L}$ we obtain the \eq{Conc301} inflation solution with the positive initial energy value of the Lorentzian space-time. Nevertheless, for the negative initial energy data, \eq{Conc4}
and \eq{Conc8} solutions, we have as well an inflating and additional collapsing space-times. In turn, for the Euclidean counterparts of the Lorentzian patch we have two different possibilities.
The first one, \eq{DW18}-\eq{DW182} solution, describe
the Euclidean space-time which behavior is fully symmetrical to the Lorentzian counterpart. In turn, there are  \eq{Conc3001},\eq{Conc7001} ans \eq{Conc9001} solutions which describe the collapsing Euclidean counterpart for the inflating 
Lorentzian patch, \eq{Conc3001} solution, and additional cases of the inflating Euclidean regions with inflation's rate different from the Lorentian's one. Interesting, that in each solution the kink's parameter $\lambda$ determines the rate of the inflation and/or collapsing. In turn, the Hubble's parameter can be positive or negative,  for the inflating or collapsing space-times, but in both cases it has a singularity at the hypersurface of separation
at $\tau\,\rightarrow\,0$. Of course it's dependence on time is obtained only at small $\tau$ and valid only till $t\propto\,\lambda^{-1}$.

 Concluding we can say that in the present article we tried to describe properties of the Euclid-Lorentz space-times transition in the framework of the simple model based on the general relativity. The interesting results obtained are that
this transition naturally describes a singularity of the initial Lorentzian manifold. The form and type of the singularity depends, of course, on the form of the $g_{00}$ metric's component chosen for the description of the transition.
It is especially simple for the case of the time's kink we considered but of course there are a plenty of possibilities exist. Another inyteresting properties of the model that as a consequence of the junction condition
on the hypersurface which separate the patches of the whole space-time, we obtain a dynamical spatial metric $h_{i j}$ for both manifolds. This tensor can be considered as transition vertex which depends on the time  and determines by the initial data for the Einstein equations. In this case we face a evolution of the manifolds which can be inflation or collapsing without any additional fields but only because of the geometry introduced and dynamics of the transition.
The obtained results, therefore, 
can be useful in the further understanding   of the dynamics of the signature change between different geometries abd clarification of the inflation properties of the dynamical geometry..

\section*{Acknowledgments}

 The authors are grateful to M.Zubkov and M.Schiffer for the useful discussions and advices concern the subject of the manuscript.

\end{document}